\documentstyle[psfig,galley]{mn}

\def\gs{\mathrel{\raise0.35ex\hbox{$\scriptstyle >$}\kern-0.6em
\lower0.40ex\hbox{{$\scriptstyle \sim$}}}}
\def\ls{\mathrel{\raise0.35ex\hbox{$\scriptstyle <$}\kern-0.6em
\lower0.40ex\hbox{{$\scriptstyle \sim$}}}}

\title[The starburst in N2\,850.4]
{A vigorous starburst in the SCUBA galaxy N2\,850.4}

\author[Smail et al.]
	{Ian\ Smail,$^{\! 1}$ S.\,C.\ Chapman,$^{\! 2}$ R.\,J.\
	Ivison,$^{\! 3}$  A.\,W.\ Blain,$^{\! 2}$ T.\ Takata,$^{\! 4}$ 
	T.\,M.\ Heckman,$^{\! 5}$ \and J.\,S.\ Dunlop$^{6}$ \& K.\ Sekiguchi$^{4}$
        \vspace*{1mm}\\
	$^1$ Institute for Computational Cosmology, University of Durham, 
	South Road, Durham DH1 3LE\\
	$^2$ California Institute of Technology, Pasadena, CA
	91125, USA\\
	$^3$ Astronomy Technology Centre, Royal Observatory, 
        Blackford Hill, Edinburgh EH9 3HJ\\
	$^4$ Subaru Telescope, National Astronomical Observatory,
	650 N. A'ohoku Place, Hilo, HI 96720, USA\\
	$^5$ Dept.\ of Physics \& Astronomy, Johns Hopkins University,
	Baltimore, MD 21218, USA\\
	$^6$ Institute for Astronomy, University of Edinburgh, 
        Blackford Hill, Edinburgh EH9 3HJ}

\date{Accepted ... ; Received February 27 2003 ; in original form November 29 2002}

\pagerange{000--000}

\begin{document}
\maketitle

\begin{abstract}
We present optical and near-infrared (IR) spectroscopy of a $z=2.38$
hyperluminous IR galaxy, covering the restframe wavelength range from
1000--5000\AA.  It appears to comprise two components separated by less
than 1$''$ on the sky ($\ls 8$\,kpc); one component (B) is blue, the
other (P) is red in restframe ultraviolet(UV)-optical colours.  The
combined system has a restframe luminosity of $\sim 8$\,L$_V^\ast$ and
its restframe optical spectrum is characteristic of a Seyfert active
galactic nucleus (AGN).  However, its restframe UV spectrum exhibits
striking features associated with young stars, including P-Cygni lines
from stellar winds and blue-shifted interstellar absorption lines
indicative of a galactic outflow.  Redshifts are derived from stellar
photospheric lines in the UV and from narrow emission lines in the
restframe optical, and these are compared to that measured for the
molecular gas recently detected with the IRAM interferometer. The offsets
indicate that the far-IR emission is most likely associated with the
near-IR source P, which hosts the Seyfert nucleus, while the UV-bright
component B is blueshifted by 400\,km\,s$^{-1}$. This suggests that the
two components are probably merging and the resulting gravitational
interactions have triggered the hyperluminous activity. Modelling of
the UV spectral features implies that the starburst within the UV
component of this system has been going on for at least $\sim
10$\,Myrs.  Assuming that the bolometrically-dominant obscured
component has a similar lifetime, we estimate that it has so far formed
a total stellar mass of $\sim 10^{11}$\,M$_\odot$.  If this star
formation continues at its present level for substantially longer, or
if this activity is repeated, then the present-day descendant of
N2\,850.4 will be a very luminous galaxy.
\end{abstract}

\begin{keywords}
cosmology: observations --- 
galaxies: evolution --- 
galaxies: formation --- 
galaxies: starburst ---
galaxies: individual (ELAIS N2\,850.4; SMM\,J16358+4057) 
\end{keywords}

\section{Introduction}

Surveys with the SCUBA and MAMBO sub/millimeter cameras have identified
a population of ultra- and hyper-luminous galaxies, L$_{\rm bol}\geq
10^{12}$ and $\geq 10^{13}$\,L$_\odot$ respectively, at high redshifts,
$z>1$ (Cowie et al.\ 2002; Dannerbauer et al.\ 2002; Scott et al.\
2002; Smail et al.\ 2002; Webb et al.\ 2002).  Unfortunately, the
modest spatial resolution of submm instruments, combined with the
relative faintness of these galaxies in optical wavebands, has meant
that detailed investigation of their properties has been limited to a
handful of cases (Ivison et al.\ 1998, 2000, 2001; Soucail et al.\
1999; Gear et al.\ 2000; Chapman et al.\ 2002a; Ledlow et al.\ 2002;
Dunlop et al.\ 2003; Smail et al.\ 2003).  However, sensitive
high-resolution radio maps provide a powerful tool for localizing the
submm emission, assuming that the tight correlation between far-IR and
radio emission in dusty galaxies continues to high redshifts.  In this
way, it has proved possible to identify counterparts to large samples
of radio-identified submm galaxies (Barger, Cowie \& Richards 2000;
Chapman et al.\ 2001; Ivison et al.\ 2002) and hence begin to make
progress on determining the redshifts of significant samples of SCUBA
galaxies (Chapman et al.\ 2003a).

To gauge the contribution of submm galaxies to the formation of the
stellar populations in galaxies at $z\sim 0$ we need to determine
whether their immense far-IR luminosities originate from
dust-reprocessed star formation or AGN activity and also what the
lifetime of this activity is.  Signatures of AGN activity have been
found in a few well-studied examples (e.g.\ Ivison et al.\ 1998; Ledlow
et al.\ 2002) and it is generally supposed that most systems will
comprise a mix of starburst- and AGN-powered emission, although the
latter may be weak (Almaini et al.\ 2002; Alexander et al.\ 2002,
2003).  Having concluded that star formation provides a significant
fraction of the bolometric luminosity from typical SCUBA galaxies, the
critical issue for understanding their contribution to the stellar
density at $z=0$ is to measure the duration of the starburst activity.
The star formation rates in representative SCUBA galaxies are very
high, of the order of 10$^3$\,M$_\odot$\,yr$^{-1}$, but if this
activity occurs in a compact nuclear burst then the activity could in
principle only last $\sim 1$\,Myrs, and thus produce only a modest mass
of stars, sufficient for a typical bulge.  Alternatively, if the burst
is more prolonged (or repeated) then much larger stellar masses could
form, providing a clear link between the SCUBA galaxies and the most
massive local stellar systems at $z=0$.  Obtaining a robust lower limit
on the lifetime of a massive starburst in a SCUBA galaxy would provide
an essential confirmation of the role of intense, obscured star
formation in the evolution of luminous elliptical galaxies.
 
Another important consequence of timing the intense activity in SCUBA
galaxies is the expectation that their activity may be
self-regulating, through the intervention of powerful winds and
outflows.  These outflows will have a profound effect on both their
internal processes (by removing the gas reservoir needed to fuel star
formation and AGN activity) and the formation and evolution of
galaxies in their immediate surroundings.  Evidence of the striking
effects of starburst- and AGN-driven winds from active galaxies at low
and high redshifts is growing (Heckman et al.\ 1990; Vernet \& Cimatti
2001; Strickland et al.\ 2002; Pettini et al.\ 2002; Ledlow et al.\
2002).  As the most luminous galaxies in the high-redshift Universe,
SCUBA sources may provide an unique laboratory for the study of the
influence of feedback on galaxy formation and the properties of the
intergalactic medium (Adelberger et al.\ 2002).

In this paper we present spectroscopic observations of a
radio-identified, submm-selected galaxy, ELAIS N2\,850.4 at $z=2.38$
(Scott et al.\ 2002; Fox et al.\ 2002; Ivison et al.\ 2002; Chapman et
al.\ 2003a), which displays the spectral signatures of a massive,
vigorous starburst. Similar features are visible in several other
galaxies in the Chapman et al.\ (2003a) sample, although these are
typically fainter galaxies and the features are therefore noisier.
Thus, by virtue of its unusual optical brightness, N2\,850.4 presents
the clearest example to date of a UV-detected starburst in a SCUBA
galaxy.  In the following we adopt a $\Omega_o=1$, $\Omega_\Lambda=0$,
H$_o=50\,$km\,s$^{-1}$\,Mpc$^{-1}$ cosmology. If we used a flat
cosmology with $\Omega_\Lambda=0.7$ and
H$_o=65\,$km\,s$^{-1}$\,Mpc$^{-1}$, then the implied luminosities will
rise by 20\,per cent and the linear scale will increase by 10\,per
cent.  We present our observations in the next section, describe our
analysis and results in \S3, discuss these in \S4 and give our
conclusions in \S5.

%
%
\begin{figure}
\centerline{\psfig{figure=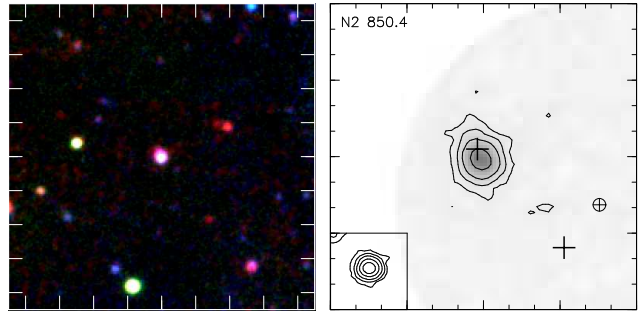,angle=0,width=3.3in}}
\caption{A $45''\times 45''$ true colour $VRK$ image of N2\,850.4
(left) showing the red emission extending out to the north of the
optical source.  Note the presence of an Extremely Red Object 11$''$ to
the West of N2\,850.4 with $K\sim 19.5$ and $I-K\sim 4.2$.  An expanded
view, $12''\times 12''$ (47\,kpc at $z=2.38$, h$_0=0.5$, q$_0=0.5$), of
the morphology of the galaxy (right) from the composite $V+R+K$ image.
The crosses show the positions of the two radio sources in the vicinity
of the submm source, whose centroid is shown by the cross-in-circle.
We only show the greyscale image within the 8$''$-radius error-circle
of the submillimetre source.  The contours are logarithmically spaced
and the image has been smoothed with a 0.4$''$ FWHM Gaussian to
suppress pixel-to-pixel noise.  To demonstrate the reality of the
extension to N2\,850.4 we show as an inset a similar contour plot of a
star 20$''$ south of the source, which shows a more strongly peaked 
and circular
morphology.}
\end{figure}

\section{Observations and Reduction}

N2\,850.4 was identified as a relatively bright submm source at
16\,36\,50.0, +40\,57\,33 (J2000), with an 850-$\mu$m flux of $8.2\pm
1.7$\,mJy and a 450-$\mu$m limit of $<$34\,mJy (3-$\sigma$), in the
survey of the ELAIS-N2 field by Scott et al.\ (2002).  Fox et al.\
(2002) provide constraints on the photometric properties of possible
counterparts lying within the large submm error circle of this source,
including 3-$\sigma$ upper-limits from {\it ISO} on the 7- and
15-$\mu$m emission of $<$1\,mJy and $<$\,2mJy respectively.  Using a
deep 1.4$''$-resolution 1.4-GHz map of the field from the VLA, Ivison
et al.\ (2002, I02) identified a compact radio source with a 1.4-GHz
flux of $221\pm 17 \mu$Jy, at 16\,36\,50.425, +40\,57\,34.46 (J2000),
within 5$''$ of the submm centroid.\footnote{A possible 3-$\sigma$
radio source lies 2.1$''$ from the nominal submm position (Fig.~1), but
based on probabilistic arguments I02 claim that this is less likely to be
the correct ID. The results presented here for the remarkable
properties of N2\,850.4 argue strongly against this association.}  A
relatively bright optical/near-IR counterpart to this radio/submm
source was also identified by I02 in their $V\!RIK$ imaging at
16\,36\,50.461, +40\,57\,34.25 (J2000) in the $R$-band and
16\,36\,50.402, +40\,57\,34.19 (J2000) in $K$: both positions are tied
to the radio reference frame with a precision of $\pm 0.3''$.  I02 list
a magnitude of $K=18.43\pm 0.02$ for N2\,850.4 and blue optical colours
$(R-I)=0.45\pm 0.03$, $(V-R)=0.12\pm 0.03$, but a relatively red
optical--near-IR colour $(I-K)=3.40\pm 0.03$.\footnote{The galactic
extinction in this field is $E(B-V)=0.006$ based on Schlegel et al.\
(1998) and so we apply no correction.} We illustrate the morphology of
N2\,850.4 as well as its wider environment, using the images from I02,
in Fig.~1.  I02 classify the system as a probable pair of galaxies,
with a blue optical source and a redder near-IR galaxy, on the basis of
small offset in the centroid in the two wavebands, $\sim 0.2''$.  This
offset is visible as the red extension to the North-East of the galaxy
in Fig.~1.  We limit the maximum possible separation of the two
components to $\ls 1''$ and measure the FWHM of the galaxy as 0.64$''$
in $V$ and 0.71$''$ in $K$, corrected for seeing.  In the following we
denote the blue optical source as component ``B'' and the redder
feature as component ``P''.

%
%
\setcounter{figure}{1}
\begin{figure*}
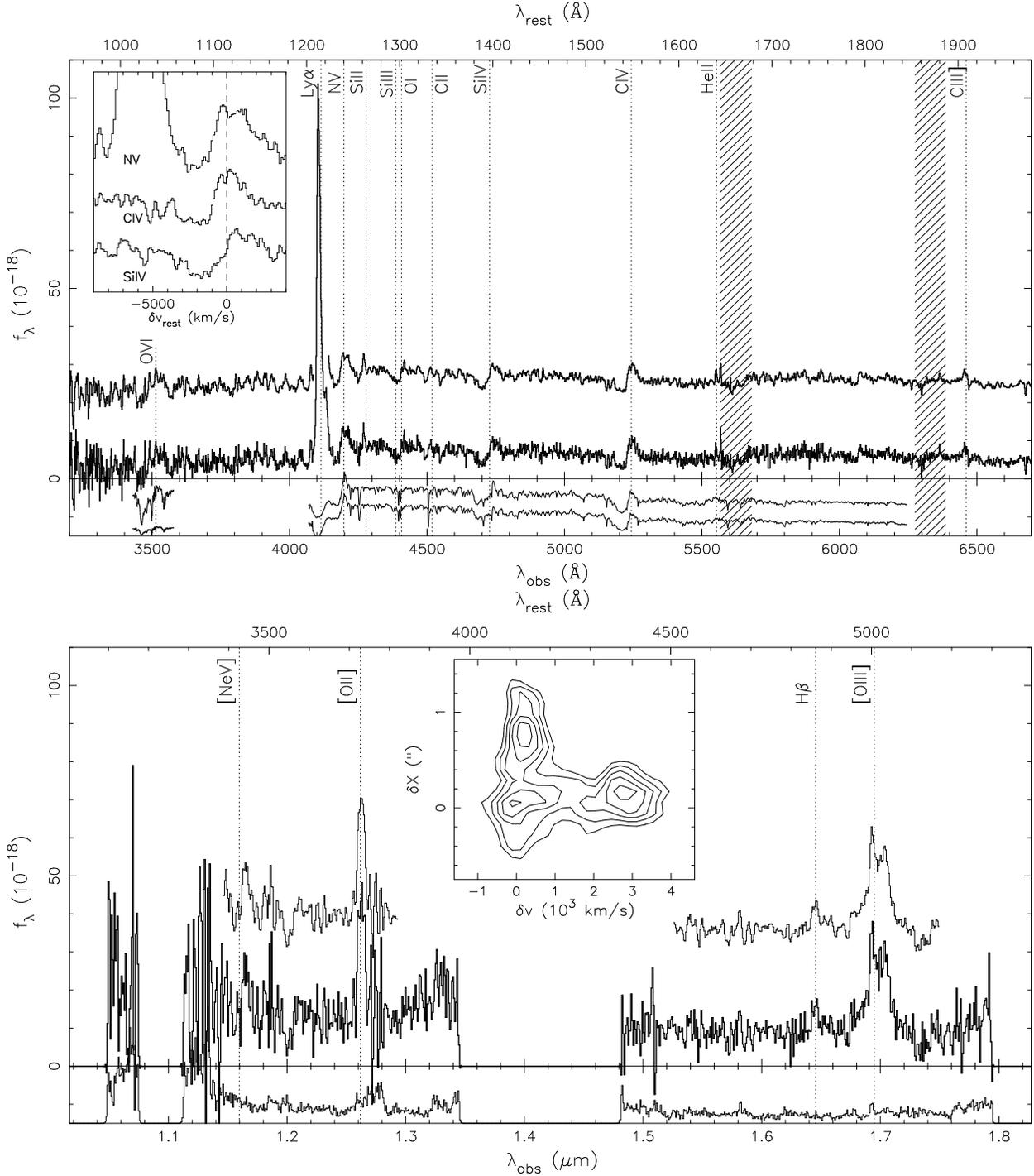

\centerline{\psfig{figure=fig2a.ps,angle=270,width=6.5in}}
\centerline{\psfig{figure=fig2b.ps,angle=270,width=6.5in}}
\caption{\small a) The upper panel is the LRIS spectrum of N2\,850.4
in the restframe UV.  The middle spectrum in this panel shows the raw
data and the upper spectrum is smoothed to the instrumental
resolution.  We mark the expected wavelengths of strong emission or
absorption features based on a redshift of the galaxy of $z=2.384$
from the observed wavelength of the CO(3--2) emission (Genzel et al.\
2003).  The hashed areas show regions of the spectrum which are
effected by strong sky emission or the dichroic in the spectrograph.
The two lower spectra in this panel show the UV spectra predicted by
the Starburst99 model assuming a 3-Myr old burst (upper) and a 10-Myr
continuous (lower) star formation models, both with a Salpeter IMF
(these have been corrected to the redshift of the photospheric
absorption lines to allow easy comparison with the blue-shifted
features in the UV).  These models illustrate the strength of the
P-Cygni features expected for stellar population dominated by young,
massive stars. The inset shows more details of the P-Cygni absorption
features seen in the LRIS spectrum.  ~~~b) The lower figure shows the
Subaru CISCO/OHS $JH$-band spectrum of N2\,850.4 covering the
restframe optical.  Again we identify the strongest emission lines in
the spectrum and note that [O{\sc ii}]\,$\lambda$3727 provides 10\,per
cent of the emission in the $J$-band and [O{\sc iii}]\,$\lambda$5007
contributes 25\,per cent of the $H$-band flux.  The upper trace is
smoothed to the instrumental resolution and the lower trace
illustrates the sky noise as a function of wavelength.  The inset
panel shows a position--velocity plot around the [O{\sc
iii}]\,$\lambda$5007 line, demonstrating the spatial and velocity
extent of this emission line.}
\end{figure*}

Spectroscopic observations of N2\,850.4 were obtained with the LRIS
spectrograph on the Keck-I telescope\footnote{The W.\,M.\ Keck
Observatory is a scientific partnership between the University of
California, the California Institute of Technology and the National
Aeronautics and Space Administration, and was made possible by the
generous financial support of the W.\ M.\ Keck Foundation.} as part of
a spectroscopic survey of radio-identified SCUBA galaxies (Chapman et
al.\ 2003a).  A total of 3.6\,ks of integration was obtained on the
night of 2002 March 19 using a slit mask with 1$''$ wide slitlets and
the 400\,l\,mm$^{-1}$ grism on the blue arm, with the
600\,l\,mm$^{-1}$ grating for the red arm.  This set-up provides a
combination of wavelength coverage down to the atmospheric cut-off in
the blue and reasonably high resolution in the red to facilitate sky
subtraction.  The PA of the mask was 80\,degrees and objects were
nodded along the slits by 2$''$ between subexposures to improve
flat-fielding and to allow the construction of a fringe frame from the
science images.  The observations were reduced in a standard manner
using {\sc iraf} tasks.  

The LRIS spectrum shown in Fig.~2a includes a series of strong emission
lines which indicate a redshift of $z\simeq 2.4$, a blue continuum
$F_\lambda \propto \lambda^{-2.1}$ and detectable emission down to
1000\AA\ in the restframe.  The identifications of some of the stronger
emission and absorption features are marked, including Ly$\alpha$,
O{\sc vi}\,$\lambda$1035, N{\sc v}\,$\lambda$1240, O{\sc
i}\,$\lambda$1302 (probably blended with Si{\sc ii}\,$\lambda$1304 and
Si{\sc iii}\,$\lambda$1295), Si{\sc iv}\,$\lambda$1396 (merged with
O{\sc iv}]\,$\lambda$1405) and C{\sc iv}\,$\lambda$1549.

The N{\sc v}\,$\lambda$1240, Si{\sc iv}\,$\lambda$1396 and C{\sc
iv}\,$\lambda$1549 lines show classic P-Cygni profiles, with
blue-shifted absorption troughs with widths of
3000--4000\,km\,s$^{-1}$, which are illustrated in more detail in the
inset panel in Fig.~2a.  Several of these lines have associated broad,
redshifted emission features.  The detailed morphologies of these
emission lines (and also Ly$\alpha$) suggest the presence of two
components to the emission: one narrow and blue, the other broader,
redshifted and spatially offset by 0.2$''$ along the slit to the
North-East.  We therefore fit the emission lines corresponding to
Ly$\alpha$, N{\sc v}\,$\lambda$1240, Si{\sc iv}\,$\lambda$1396+O{\sc
iv}]\,$\lambda$1405 and C{\sc iv}\,$\lambda$1549, with two Gaussian
components (Table~1).

In addition to the strong absorption features from stellar winds we
identify interstellar absorption lines from Si{\sc ii}\,$\lambda$1260
and C{\sc ii}\,$\lambda$1335 at $z=2.379\pm 0.002$ and $z=2.373\pm
0.002$ respectively.  We also see weak emission features associated
with fine-structure lines of Si{\sc ii} at 1265\AA\ and 1309\AA\
(Table~1).  Finally, there are a number of weak, narrow photospheric
absorption lines of S{\sc v}\,$\lambda$1502, C{\sc iii}\,$\lambda$1427
and Si{\sc iii}\,$\lambda$1417 visible in the spectrum, with restframe
equivalent widths 0.5--1\AA; we use these below to estimate the
systemic redshift of the stars dominating the UV emission.

%
%
\begin{table*}
 \centering
 \begin{minipage}{130mm}
 \caption{Properties of restframe optical and UV emission lines of N2\,850.4}
\begin{tabular}{lcccrcl}
\noalign{\hrule \smallskip}
Line      & $\lambda_{\rm rest}$ & $\lambda_{\rm obs}$ & $z$ &  Width~ & Flux & Comment \cr
          & (\AA)               & (\AA)               &     &
(km\,s$^{-1}$) &  (erg\,s\,cm$^{-2}$)  & \cr
\noalign{\smallskip \hrule \smallskip}
Ly$\alpha$ & 1215 & 4111.4 & 2.382  & 900 & $6.9\times 10^{-16}$ & Narrow component \cr
          &  & 4126.7 & 2.395  & 2600 & $1.2\times 10^{-15}$ &  Broad component \cr
N{\sc v}   & 1239 & 4199.1 & 2.386  & 500 & $2.8\times 10^{-17}$ &   Narrow component \cr
           &      & 4214.1 & 2.398 & 2500 &$1.9\times 10^{-16}$ & Broad component\cr
Si{\sc ii} & 1265 & 4269.6 & 2.376  & 560 & $4.0\times 10^{-17}$  & \cr
O{\sc i}   & 1302 & 4423.0 & 2.394  & 700 & $1.7\times 10^{-17}$  &
Blended with Si{\sc iii}\,$\lambda$1295 and Si{\sc ii}\,$\lambda$1304
\cr
Si{\sc ii} & 1309 & 4416.3 & 2.373  & 300 & $1.2\times 10^{-17}$  & \cr
Si{\sc iv}+O{\sc iv}] & 1404 & 4742.5 & 2.378 & 400 & $4.7\times 10^{-17}$  & Narrow component \cr
           &  & 4761.9 & 2.392 & 800 & $6.9\times 10^{-17}$  & Broad component \cr
C{\sc iv}  & 1549 & 5241.0 & 2.383 &  500 & $1.7\times 10^{-17}$  & Narrow component \cr
           &  & 5256.6 & 2.394 &  1600 & $1.3\times 10^{-16}$  & Broad component \cr
He{\sc ii} & 1640 & 5542.0 & 2.378 &  $\leq 300$ & $2.0\times 10^{-17}$ & \cr
C{\sc iii}] & 1909 & 6461.6 & 2.385 & 850 & $3.7\times 10^{-17}$  & \cr
\noalign{\smallskip}
[Ne{\sc v}] & 3426 & 11642 & 2.400 &  $<$1500 & $7.4\times 10^{-17}$ &  \cr
[O{\sc ii}] & 3727 & 12624 & 2.387 &  $<$1500 & $2.4\times 10^{-16}$ &  \cr
H$\beta$    & 4861 & 16447 & 2.383 & $<$1500 & $2.6\times 10^{-17}$ &  Narrow component  \cr
    &  &  16495  & 2.393 & 4000  & $5.8\times 10^{-17}$  & Broad component  \cr
[O{\sc iii}] & 5007 & 16924 & 2.380 &   $<$1500 & $2.1\times 10^{-16}$  & Spatially extended \cr
          &  & 17030 & 2.401 &  1900 & $2.8\times 10^{-16}$ & Red component \cr
\noalign{\smallskip \hrule} 
\end{tabular}
\end{minipage}
\end{table*}
\medskip

%
%
\begin{table*}
 \centering
 \begin{minipage}{130mm}
 \caption{Suggested groupings of different redshift components and
probable origins in N2\,850.4.}
\begin{tabular}{lcrll}
\noalign{\hrule\smallskip} 
~~~~~~~Lines & $z$ & $\Delta v$~~~ & Component & Comment \cr
   &     &  (km\,s$^{-1}$) & & \cr
\noalign{\smallskip \hrule \smallskip}
CO emission                   & 2.384$\pm$0.001 & 0 & & Genzel et al.\
(2003) \cr
\noalign{\smallskip}
Narrow optical emission       & 2.384$\pm$0.003 & 0 & P -- Systemic  & [O{\sc ii}], H$\beta$, [O{\sc iii}] \cr
\noalign{\smallskip}
UV photospheric absorption   & 2.380$\pm$0.002 & $-$350 & B -- Systemic  &  \cr
UV interstellar absorption & 2.376$\pm$0.003 & $-$700 & B -- Outflow  & Si{\sc ii},
C{\sc ii} \cr
Narrow UV emission      & 2.382$\pm$0.003 & $-$150 & B -- Outflow & Ly$\alpha$, N{\sc v}, Si{\sc iv}+O{\sc iv}, C{\sc iv} \cr
\noalign{\smallskip}
Broad UV emission          & 2.395$\pm$0.003 & 1000 & P -- Outflow  & Ly$\alpha$, N{\sc v}, Si{\sc iv}+O{\sc iv}, C{\sc iv} \cr
Broad optical emission   & 2.401$\pm$0.001 & 1500 & P -- Outflow & [Ne{\sc v}], broad H$\beta$, broad [O{\sc iii}] \cr
\noalign{\smallskip \hrule}
\end{tabular}
\end{minipage}
\end{table*}
\medskip

To investigate the spectral properties of N2\,850.4 in the restframe
optical we also obtained a $JH$-band spectrum of this galaxy on the
night of 2002 May 18, using the near-IR spectrograph, CISCO, and the
OHS atmospheric suppression unit on the Subaru
telescope\footnote{Based on data collected at Subaru Telescope, which
is operated by the National Astronomical Observatory of Japan.}
(Iwamuro et al.\ 2001).  A total of 7\,ks of data was acquired in
0.4--$0.6''$ seeing as a series of 1\,ks exposures. We employed a
1$''$-wide slit at a PA of 0\,degrees and nodded the source along the
slit by $10''$ between exposures.  Atmospheric transmission
corrections came from SAO standard stars and wavelength calibration
was obtained from the OH lines measured in observations without OHS.
Checks of the instrument stability suggest that this calibration is
good to $<7$\,\AA.  The data were reduced in a
standard manner, including flat-fielding, sky subtraction, correction
of bad pixels and residual sky subtraction. The final $JH$ spectrum is
shown in Fig.~2b, where we identify several strong lines, including
[O{\sc ii}]\,$\lambda$3727, H$\beta$ and [O{\sc iii}]\,$\lambda$5007.
The details of the emission line measurements from the CISCO/OHS
spectrum are given in Table~1.  There are two velocity components in
the H$\beta$ and most strikingly in the [O{\sc iii}]\,$\lambda$5007
line: we list the double Gaussian fits to both lines in Table~1. The
limits on the [O{\sc ii}]\,$\lambda$4959 emission corresponding to the
[O{\sc iii}]\,$\lambda$5007 components are consistent with the
standard flux ratio of these lines.  Note that the lower redshift
component of the [O{\sc iii}]\,$\lambda$5007 line is extended along
the slit to the North by $\ls 1.5''$ and to the South by $\sim
0.5''$(see the inset in Fig.~2b).  A spectrum of the extended emission
shows not only [O{\sc iii}]\,$\lambda$5007, but also [O{\sc
ii}]\,$\lambda$3727; narrow H$\beta$ and [O{\sc iii}]\,$\lambda$4959, 
components suggest in situ ionization of the gas.  In
all cases the flux or equivalent width quoted in Table~1 is for the
emission from the core of the galaxy.

\section{Analysis and Results}

As Table~1 demonstrates, there are a bewildering range of different
spatial and velocity components in the spectra of N2\,850.4, with
redshifts spanning $z=2.373$--2.401.  To bring some coherence to the
picture we suggest a number of groupings for these components in
Table~2. We also add two further redshift estimates.  The first comes
from cross-correlating the weak photospheric lines in the UV spectrum
with template models from Starburst99 (after masking all the
interstellar and wind features).  This provides our best estimate of
the systemic redshift of the young UV stellar population: $z=2.380$,
which is very close to the redshift of the narrow Ly$\alpha$
component.  The second measurement comes from recent observations of
CO(3--2) molecular emission using the IRAM Plateau de Bure
interferometer (Genzel et al.\ 2003), which yields $z=2.384$ for the
systemic redshift of the large mass of molecular gas in this galaxy,
and should provide the most reliable estimate of the redshift of the
luminous submm source.

The redshift of $z=2.384$ for N2\,850.4 is higher than that inferred by
I02 from the radio/submm spectral index of the source (see also Fox et
al.\ 2002), $z=1.3^{+0.6}_{-0.4}$ based on the Carilli \& Yun (2000)
calibration. It is just within the 1-$\sigma$ bound of the Rengarajan
\& Takeuchi (2001) version of the indicator, which yields
$z=1.8^{+0.6}_{-0.4}$.  This suggests either an additional contribution
from an AGN to the radio emission, or that the dust emission from
N2\,850.4 is characterized by a particularly hot dust temperature
(Blain 1999; Blain, Barnard \& Chapman 2003; Chapman et al.\
2003b). Using a hotter range of dusty galaxy SEDs, Aretxaga et al.\
(2002) estimate $z=2.9^{+1.2}_{-0.9}$ for N2\,850.4 from its 1.4-GHz,
1.3-mm and 850-$\mu$m detections and a 450-$\mu$m upper limit.  The
allowed redshift range is relatively broad, but does include the true
redshift.  It is interesting to consider the influence of a short
duration for the star formation on the radio emission from N2\,850.4:
if the starburst is extremely young, then no supernovae will yet have
occurred, perhaps suppressing its radio emission, and reducing its
estimated redshift using the Carilli--Yun indicator.

One further complication in relating the UV and far-IR properties of
N2\,850.4 is our lack of knowledge about the {\it exact} location of
the sites of emission in the two wavebands. The precision of the
relative optical-radio astrometry from I02 is $\pm 0.3''$, similar to
the offset (to the North) of the radio source relative to the optical
counterpart (Fig.~1), which is therefore not significant.  As
discussed by Ivison et al.\ (2001), the complex mix of emission and
dust absorption in SCUBA galaxies could result in our sampling very
different regions of these systems at different wavelengths (see
Goldader et al.\ 2002 reach a similar conclusion based on UV
observations of local ULIRGs).  Indeed, colour differences within
N2\,850.4 (see Fig.~1; I02) suggest that dust obscuration in this
galaxy may be highly structured.  Alternatively, the colour
differences may originate from spatially distinct components within
the system (I02).

The proposed groupings of the various velocity components in N2\,850.4
(Table~2) is motivated by the likely origin of the different spectral
features and their spatial distribution.  We identify two key
components. The first dominates the UV luminosity of the system and
contributes narrow UV emission lines and weak photospheric absorption
features at a redshift $z\simeq 2.381$, this we equate with component
B.  The second component produces a number of narrow restframe optical
emission lines in the OHS spectrum at a redshift close to that measured
for the molecular gas, $z\simeq 2.384$ ($400\pm200$\,km\,s$^{-1}$
offset from the UV source), and most likely arises in the near-IR
component, P.  We therefore associate the near-IR emission with the
source of the extremely luminous, dust-reprocessed far-IR emission.

Having isolated the features from the two putative components of
N2\,850.4, we can study their spectral properties in more detail.  We
use the restframe optical emission line properties to classify
N2\,850.4 spectrally.  Several properties suggests that N2\,850.4/P
harbours an AGN: the redshifted, broad emission lines in the restframe
optical (and UV) and the apparent detection of the high ionisation
[Ne{\sc v}]\,$\lambda$3426 line, indicating hard radiation from an AGN.
The flux ratio of [O{\sc iii}]\,$\lambda$5007 to H$\beta$ is $\sim 5.8$
(including flux from both broad and narrow [O{\sc iii}] and H$\beta$
lines), which suggests a Seyfert-2 classification, while the presence
of broad H$\beta$ would instead argue for the AGN component lying in the
Seyfert-1 class. Hence, N2\,850.4/P appears to be a transition object
with characteristics of both Seyfert-1's and 2's, although the very broad
[O{\sc iii}]\,$\lambda$5007 line mean that it doesn't fit easily into
any of the existing transition classes (Osterbrock 1981).

AGN features are most obvious in the restframe optical spectrum, but
they also appear in the restframe UV (where they are spatially offset
from the UV continuum emission). In particular, the strength and
breadth of the Ly$\alpha$, N{\sc v}\,$\lambda$1239, Si{\sc
iv}\,$\lambda$1397 and C{\sc iv}\,$\lambda$1549 emission lines require
a broad AGN contribution. These features suggest only a modest AGN
contribution to the restframe UV emission, for example the C{\sc
iii}]\,$\lambda$1909 line is weaker than the C{\sc iv}\,$\lambda$1549
line.  Indeed, the most striking features in the UV spectrum are the
broad, blue-shifted absorption troughs in the blue wings of the N{\sc
v}\,$\lambda$1239, Si{\sc iv}\,$\lambda$1397 and C{\sc
iv}\,$\lambda$1549 lines (see the inset in Fig.~2a). In principle,
these features could be generated by any source of outflowing gas with
the correct physical conditions; however, a stellar origin is
indicated by the close agreement between the strengths and kinematics
of the lines and the predictions for a population of hot stars in
models from the Starburst99 library (Leitherer et al.\ 1999). Fitting
these models either to the whole spectral range from 1230--1630\AA\ or
just to spectral windows centred on the main P-Cygni lines yields
similar results: the UV spectral features are reproduced only by a
stellar population dominated by massive, young O and B stars.  We show
two examples of acceptable Starburst99 models in Fig.~2a: a 3-Myr old
starburst with a Salpeter IMF up to 100\,M$_\odot$ and $E(B-V)=0.13\pm
0.02$ based on the parameterized dust model from Calzetti et al.\
(2000), and a model with continuous star formation, an age of 10 Myrs,
and the same IMF and extinction law.  The broad agreement between the
strengths and velocity widths of N{\sc v}\,$\lambda$1239, Si{\sc
iv}\,$\lambda$1397 and C{\sc iv}\,$\lambda$1549 troughs is
excellent. Note that the strength of the Si{\sc iv}\,$\lambda$1400
wind feature supports the younger estimate, but the modest
signal-to-noise of the spectrum precludes a stronger conclusion.  We
conclude that the restframe UV emission from N2\,850.4/B is dominated by
light from a young, intense starburst, which is probably of the order
of 10\,Myrs old.

The three groups of spectral features associated with N2\,850.4/B show
similar behaviour to that seen in Lyman-break Galaxies (Shapley et al.\
2003), with the redshift of the UV photospheric absorption lying
approximately mid-way between the narrow Ly$\alpha$ redshift and the
interstellar lines.  Quantitatively, if we adopt a systemic velocity of
the UV-bright starburst of $z=2.381$ from the photospheric absorption
lines, then the redshifts for the two strong interstellar absorption
lines visible in Fig.~2a: Si{\sc ii}\,$\lambda$1260 and C{\sc
ii}\,$\lambda$1335 at $z=2.376$ suggests blue shifts of the order of
$400\pm 300$\,km\,s$^{-1}$ and comparable line widths. If this material
actually originated in the more obscured component of this system then
the relative velocity of the material is closer to
$-800$\,km\,s$^{-1}$.  In either case, the blue shift is an unambiguous
signature of the presence of outflowing material, probably resulting
from a galactic wind driven by the vigorous starburst activity (e.g.\
Heckman et al.\ 1998; Pettini et al.\ 2002).  The large velocity
offsets seen in the high excitation restframe optical lines indicate
that material is also outflowing on a smaller scale around the AGN.

In summary, we propose that N2\,850.4 comprises two related, but
independent, components, a highly obscured, gas-rich galaxy with a
Seyfert nucleus, N2\,850.4/P, and a young, modestly obscured UV-bright
starburst companion, N2\,850.4/B.  These two components are offset
spatially by $\ls 1''$ and kinematically by $\sim 400$\,km\,s$^{-1}$.

\section{Discussion}

To quantify the extent of the star formation activity in N2\,850.4 we
look at its far-IR and optical luminosities. At $z=2.4$ N2\,850.4 is a
hyperluminous galaxy with a bolometric luminosity of $(3\pm 2) \times
10^{13}$\,L$_\odot$, based on its submm flux and a dust temperature of
$T_{\rm d}=58$\,K derived from the far-IR--radio correlation (Chapman
et al.\ 2003a).  If this emission arises purely from massive star
formation, then the luminosity corresponds to a star formation rate of
$\sim6\times 10^3$\,M$_\odot$\,yr$^{-1}$ for stars more massive than
1\,M$_\odot$ (extrapolating down to 0.1\,M$_\odot$ the rate would rise
by a factor of 3.2).  However, a significant AGN contribution to the
radio emission would lower the derived dust temperature, and thus both
the bolometric luminosity and the implied star formation rate.  For
example, if we adopt an extreme model where 75\,per cent of the radio
emission is generated directly by the AGN, then the best-fit dust
temperature drops to $T_{\rm d}=31$\,K and the bolometric luminosity
declines by a factor of $\sim 10$.

To improve the precision of the luminosity and dust temperature
measurements we require mid- and far-IR photometry from {\it SIRTF}.
Using our current best-fit SED we predict fluxes for N2\,850.4 in the
{\it SIRTF} bands at 24, 70 and 160\,$\mu$m of 3.3, 25 and 110\,mJy
respectively.  These fluxes are factors of 2--8 times greater than the
expected confusion limits in these bands and hence the source should
be easily detectable in the SWIRE survey of this region. 
These observations should provide a conclusive measurement of the
temperature of the dust in the galaxy and a test of the form of the
far-IR--radio correlation at high-$z$.

Several of the properties of N2\,850.4 outlined in the previous
sections are reminiscent of other well-studied SCUBA galaxies, many of 
which contain a partially-obscured and relatively modest
luminosity AGN (Ivison et al.\ 1998; Soucail et al.\ 1999; Ledlow et
al.\ 2002; Smail et al.\ 2003; Chapman et al.\ 2003; Alexander et al.\
2003).  What is remarkable about N2\,850.4 are the strong signatures
of vigorous on-going star formation, especially the P-Cygni absorption
troughs.  This combination of features is not unknown in galaxies at
$z\sim 0$: a local analogue of the spectrum of N2\,850.4 is provided
by the Seyfert-2 galaxy NGC\,7130 (Gonz\'alez Delgado et al.\ 1998;
Contini et al.\ 2002).  

The {\it Hubble Space Telescope} GHRS spectrum of NGC\,7130 (Fig.~8 in
Gonz\'alez Delgado et al.\ 1998) is strikingly similar to that of
N2\,850.4: both galaxies show weak emission lines of N{\sc v}, S{\sc
iv} and C{\sc iv}, all with strong P-Cygni absorption
troughs. Gonz\'alez Delgado et al.\ interpret these spectral features
as the result of a nuclear starburst with an age of 3--4\,Myr.
NGC\,7130 also shows two components of [O{\sc iii}]\,$\lambda$5007
with very different widths. The narrower line is blueshifted with
respect to the broader component that results from emission in
outflowing gas driven by an AGN or starburst. In N2\,850.4 we see a
much larger velocity shift between the two components.  The narrower,
blue component is spatially extended along our slit and exhibits a
small velocity shear of order 160\,km\,s$^{-1}$ on an arcsec scale.
The most significant difference between NGC\,7130 and N2\,850.4 is
their luminosities: at $3\times 10^{11}$L$_\odot$ NGC\,7130 is two
orders of magnitude less luminous than N2\,850.4.

N2\,850.4 is also astonishingly bright in the restframe optical
waveband.  The observed $K$-band magnitude of $K=18.43$ (I02) implies
a restframe $V$-band absolute magnitude of M$_V\simeq -24.8$.
However, its red $(I-K)$ colour (I02) may reflect a strong
contribution from the H$\alpha$ line to the $K$-band flux.  Assuming
that 50\,per cent of the $K$-band light comes from the H$\alpha$ line
(similar to the emission line contribution in the $H$-band), this
would reduce the absolute magnitude to M$_V\sim -24.0$, but still
indicating a very luminous galaxy.  

In view of the extreme luminosity of N2\,850.4 we
should consider the possibility that strong gravitational lensing
artificially boosts its apparent brightness, as seen in several
extremely luminous high-$z$ sources, including some SCUBA galaxies
(e.g.\ Graham \& Lui 1995; Williams \& Lewis 1996; Ibata et al.\ 1999;
Solomon \& Downes 2002; Chapman et al.\ 2002b; Dunlop et al.\ 2003).
The apparent offset between the optical and near-IR emission from
N2\,850.4 could be due to the offset between the source and a
foreground galaxy lens.  The spectral properties of N2\,850.4 suggest
that if a lens is present, then it contributes neither much light to
the optical spectrum shortward of $\sim 6000$\AA\ nor any emission
lines in the optical/near-IR windows.  The most obvious class of lens
is thus an early-type galaxy at $z\sim 0.5$--1. However, only a
sub-$L^*$ lens would not exceed the $K$-band magnitude, implying an
Einstein radius of only $\sim 0.1''$ in the absence of any local large
scale structure. For significant strong lensing amplification ($\gg
10\times$), we would expect the lens and source to be aligned at the
$\sim 0.1''$ level, similar to the observed offset in the centroids in
the optical and near-IR wavebands (Ivison et al.\ 2002).  However, the
detection of spatially-extended optical and near-IR line emission
argues against significant lens amplification, which would imprint a
strong amplification gradient on this emission. We do not believe that
N2\,850.4 is strongly affected by gravitational lensing. However, a
definitive conclusion must await either higher resolution imaging, or
preferably, spatially-resolved 2-dimensional spectroscopy.

The distinct optical and near-IR components in the system are separated
by less than 8\,kpc in projection and the apparent velocity offset
between their restframe UV photospheric and narrow optical emission
line redshifts is $\sim 400$\,km\,s$^{-1}$. If the two components are
bound/merging, then the total mass of the system must be around
$1\times 10^{12}$\,M$_\odot$, which is comparable to previous dynamical
mass estimates for SCUBA galaxies based on CO line widths (e.g.\ Frayer
et al.\ 1998, 1999; Genzel et al.\ 2002; Solomon \& Downes 2002).
While the velocity offset is uncertain, it is consistent with the
identification of N2\,850.4 as a massive galaxy undergoing an
interaction with a companion.

Assuming a velocity dispersion for N2\,850.4 comparable to the
velocity offset between the optical and near-IR components, $\sim
400$\,km\,s$^{-1}$, then the crossing time for the galaxy is $\sim
10$\,Myrs, similar (given the large uncertainties) to the estimated
age of the starburst and supporting the interpretation of the UV
starburst as an instantaneous event triggered by the interaction.
This suggests that the same timescale should be applied to the
obscured activity which is powering the far-IR emission.

The near-IR luminosity of N2\,850.4 implies a very bright absolute
magnitude $M_V < -24$ (assuming it is not strongly lensed), which
corresponds to a $3\times 10^{9}$M$_\odot$ starburst for a Salpeter IMF
from 1--100\,M$_\odot$ for continuous star formation in the 10-Myr
Starburst99 model with no dust reddening.  Assuming a constant star
formation rate over the lifetime of the burst then suggests that the
galaxy must have been forming stars at a rate of $\sim 3\times
10^2$\,M$_\odot$\,yr$^{-1}$.  Although there is a large uncertainty in
this estimate, it is far below the star formation rate determined from
the far-IR luminosity.  The predicted bolometric luminosity of the
optical component from the Starburst99 model is $\sim
10^{12}$\,L$_\odot$, again roughly 10\,per cent of that estimated from
the 850-$\mu$m and radio flux.  Any reddening to the population
dominating the restframe optical emission would clearly increase this
estimate. Using the extinction inferred from the restframe UV spectrum
gives $A_V\sim 0.4$, or a correction of 50\,per cent.  Alternatively,
using the empirical relationships derived by Meurer, Heckman \&
Calzetti (1999) between the spectral slope at 1600\AA\ and the ratio of
far-IR to 1600\AA\ luminosities for local lower-luminosity starbursts,
we estimate a similar factor of ten shortfall between the predicted
bolometric luminosity of N2\,850.4 based on its restframe UV flux and
that inferred from submm and radio observations.

The order of magnitude difference in the bolometric luminosity
estimated from the young stellar population in the UV and that
detected in the submm waveband is consistent with these arising in
distinct components.  This supports our earlier proposal that N2\,850.4
consists of a lightly-obscured young stellar population providing
$\sim 10$\,per cent of the bolometric luminosity (B), and a highly-obscured
starburst/AGN (P) which produces the bulk of the remainder.  Hence, in the
following we assume that the starburst activity seen in the UV and the
submm have separate origins. However, we caution that the only way to
confirm this hypothesis is to obtain higher resolution observations of
N2\,850.4 at both wavelengths.

Turning to the UV starburst in N2\,850.4/B, we can use the size of the
starburst and the star formation rate to estimate the star formation
density in this component.  As the restframe UV emission from the
galaxy is dominated by light from very young stars, we estimate the
physical size of the starburst from the $V$-band (restframe 1600\AA)
extent of the galaxy: 0.64$''$, or 5.0\,kpc.  Interestingly, the large
extent of the starburst region is also supported by the apparently
modest wind velocity given the strength of the UV starburst.  Using
the analytic wind model of Shu, Mo \& Mao (2003), a star formation
rate of 300\,M$_\odot$yr$^{-1}$ and a wind velocity of $\sim
350$\,km\,s$^{-1}$ we predict that the starburst region should be
$\sim 10$\,kpc in diameter.  Moreover, this same model predicts that
the wind should have substantial mass-loading, $\stackrel{\rm
.}{M}\sim10^3$M$_\odot$yr$^{-1}$.

Using the estimated size of the starburst and adopting the
extinction-corrected star formation rate determined from the UV flux
above, this yields a star formation density of $\sim
25$\,M$_\odot$\,kpc$^{-2}$\,yr$^{-1}$, with an uncertainty of a factor
of at least two.  This is similar to the claimed limiting intensity
for the star formation density in local UV starbursts derived by
Meurer et al.\ (1997), 45\,M$_\odot$\,kpc$^{-2}$\,yr$^{-1}$.  If the
UV component of N2\,850.4 is forming stars at a maximal rate, then
this suggests that the same physical mechanisms which occur at low
redshift to limit starbursts are also operating in this luminous
system at $z=2.38$.  This should give some encouragement that feedback
mechanisms calibrated at low redshift may be applicable at these much
earlier times.  Trying to apply this approach to the obscured
component will require better constraints on the extent of the far-IR
emission in this galaxy and the contribution from an AGN to the radio
and far-IR emission.

Given the enormous star formation rate of the burst, is this truly a
primeval galaxy undergoing its first star formation event?  We believe
that this is unlikely as the optical emission features we see suggest
an AGN is present in the system, and this requires that a relatively
massive black hole has had time to grow.  Using the [O{\sc
iii}]\,$\lambda$5007 luminosity of the galaxy, L$_{\rm [O{\sc iii}]} =
1.7\times 10^{42}$\,erg\,s$^{-1}$, then the correlations in
Alonso-Herrero et al.\ (1997, 2002) suggest a black hole mass of $\gs
10^8$\,M$_\odot$.  Assuming that black holes grow through accretion
with a radiative efficiency of $\sim 0.1$ then such a massive black
hole would take $\gs 100$\,Myrs to grow even at a constant accretion
luminosity of order $10^{13}$\,L$_\odot$ -- much longer than the
estimated timescale of the current burst -- suggesting that this is
unlikely to be the first, significant star formation event in this
galaxy.  Thus if the star formation and AGN activity in young galaxies
are related, then both types of activity may be intermittent (Page et
al.\ 2001; Almaini et al.\ 2002).  Nevertheless, the current burst is
still capable of producing very large numbers of stars: if we assume
the lifetime of the obscured starburst is the same as that in the UV
companion, 10\,Myrs, then the star formation rate implies that up to
$\sim 10^{11}$\,M$_\odot$ may have already been formed.

We note that N2\,850.4 is the only one of $\ls 10$ SCUBA galaxies
(with UV spectroscopy sufficiently deep to identify the spectral
signatures of a young starburst, $\ls 10$\,Myr) which show obvious
starburst features (Ivison et al.\ 2000; Chapman et al.\ 2003a).
Assuming that all SCUBA galaxies follow the same evolutionary path,
with a UV-detectable phase for 10\,per cent the submm-luminous phase
would then have a timescale of order 100\,Myr -- allowing the systems
to form 10$^{12}$\,M$_\odot$ of stars.  In that case, we are seeing
the formation of very massive galaxies. Alternatively, SCUBA galaxies
may typically not appear in a UV-luminous phase like N2\,850.4, or
their young starburst features may be weaker, and masked by
featureless UV emission from an AGN.

\section{Conclusions}

Our restframe optical and UV spectroscopy of N2\,850.4 reveal a
multitude of features associated with starburst activity, including an
array of interstellar absorption lines and P-Cygni wind absorption
profiles, which confirm the stellar origin of the continuum emission in
the restframe UV.  We also identify the spectral signatures of
large-scale gas outflows, which may be driving significant amounts of
enriched material into the galaxy's halo and the surrounding
intergalactic medium, and which may eventually remove enough gas to
halt the starburst.  The inferred luminosity of the young stellar
population requires a star formation rate of $\sim
300$\,M$_\odot$\,yr$^{-1}$ in the UV-bright component. More
importantly, the strength of the spectral features is consistent with a
young ($\gs 10$\,Myr old) starburst. Such obvious signs of a young
starburst have not previously been observed in the SCUBA population.

However, from spectroscopic data the UV bright component, N2\,850.4/B,
appears to be spatially and kinematically distinct from the near-IR
component of the system, N2\,850.4/P, which has a similar redshift to
the molecular gas reservoir recently detected by Genzel et al.\ (2003),
indicating it is likely to be the source of the luminous far-IR
emission. The interaction between the two components probably triggered
the hyperluminous far-IR activity.  The velocity offset between the
components is $\sim 400$\,km\,s$^{-1}$, suggests that, if the UV bright
companion is in a bound orbit, N2\,850.4 is a very massive galaxy,
$\sim 10^{12}$\,M$_\odot$.

As with other well-studied SCUBA galaxies, N2\,850.4 shows spectral
features of a partially obscured AGN, which we locate in the far-IR
luminous component of this system, N2\,850.4/P.  Nevertheless, the
modest line widths and luminosity of the AGN suggest it is unlikely to
dominate the energy output of the whole system and so we conclude that
the bulk of bolometric emission from N2\,850.4 is derived from
reprocessed radiation from highly-obscured, young stars.

The most interesting result that can be derived from our spectra is the
constraint on the minimum timescale of the unobscured activity in this
system, of the order of $\gs 10$\,Myr.  Assuming that the far-IR
emission was triggered at the same time, then the obscured starburst is
also at least 10\,Myrs old, and from the bolometric far-IR luminosity,
we can infer that approximately $10^{11}$\,M$_\odot$ of stars have been
formed.

However, we note that the likely timescale for the formation of the
massive black hole in the near-/far-IR luminous system is significantly
longer than $\sim 10$\,Myr, implying both that this is probably not the
first starburst experienced by this galaxy, and that star-formation
activity in the SCUBA population may occur in brief, repeated bursts,
triggered by a series of interactions and mergers. Although brief,
these bursts are still powerful: in the current starburst this massive
system could have formed an entire L$^\ast$ galaxy's worth of
stars. This underlines the possible association of SCUBA galaxies with
the formation phase of the most luminous galaxies seen in the local
Universe: giant ellipticals.

\section*{acknowledgements}

We thank Christy Tremonti for deriving the photospheric redshift for
N2\,850.4.  We thank Omar Almaini and Chris Willott for permission to
use their imaging of N2\,850.4 and Reinhard Genzel, Roberto Neri, Frank
Bertoldi, Pierre Cox and Alain Omont for allowing us to use information
from our joint PdB programme prior to publication.  We also thank the
referee for a concise and constructive report. We acknowledge useful
conversations with Alastair Edge, Cedric Lacey, Claus Leitherer, Max
Pettini, Bianca Poggianti, Alice Shapley and Chuck Steidel. IRS
acknowledges support from the Royal Society and the Leverhulme Trust.


\begin{thebibliography}{99}

\bibitem{} Adelberger, K., Steidel, C.C., Shapley, A.E., Pettini, M.,
2002, ApJ, in press.

\bibitem{} Almaini, O., et al., 2002, MNRAS, in press.

\bibitem{} Alexander, D.,  et al., 2003, AJ, in press.

\bibitem{} Alexander, D.M., Aussel, H., Bauer, F.E., Brandt, W.N.,
Hornschemeier, A.E., Vignali, C., Garmire, G.P., Schneider, D.P., 2002,
586, L85.

\bibitem{} Alonso-Herrero, A., Ward, M.J., Kotilainen, J.K., 1997,
MNRAS, 288, 977.

\bibitem{} Alonso-Herrero, A., Ivanov, V.D., Jayawardhana, R.,
Hosokawa, T., 2002, ApJ, 571, L1.

\bibitem{} Aretxaga, I., Hughes, D.H., Chapin, E.L., Gaztanaga, E.,
Dunlop, J.S., 2002, MNRAS, submitted.

\bibitem{} Barger, A.J., Cowie, L.L., Richards, E.A., 2000, AJ, 119, 2092.

\bibitem{} Blain, A.W., 1999, MNRAS, 309, 955 

\bibitem{} Blain, A.W., Barnard, V.\,E., Chapman, S.\,C., 2003, 
MNRAS, in press (astro-ph/0209450)

\bibitem{} Calzetti, D., Armus, L., Bohlin, R.C., Kinney, A.L.,
Koornneff, J., Storchi-Bergmann, T., 2000, ApJ, 533, 682.

\bibitem{} Carilli, C.L., Yun, M.S., 2000, ApJ, 539, 1024.

\bibitem{} Chapman, S.C., Richards, E.A., Lewis, G.F., Wilson, G.,
Barger, A.J., 2001, ApJ, 548, L147.

\bibitem{} Chapman, S.C., Shapley, A., Steidel, C., Windhorst, R.,
2002a, 572, L1.

\bibitem{} Chapman, S.C., Smail, I., Ivison, R.J., Blain, A.W., 2002b,
MNRAS, 335, L17.

\bibitem{} Chapman, S.C., Blain, A.W., Ivison, R.J., Smail, I., 2003a,
Nature, submitted.

\bibitem{} Chapman, S.C., et al., 2003b,
ApJ, submitted.

\bibitem{} Contini, M., Radovich, M., Rafanelli, P., Ritcher, G.M.,
2002, ApJ, 572, 124.

\bibitem{} Cowie, L.L., Barger, A.J., Kneib, J.-P., 2002, AJ, 123, 2197.

\bibitem{} Dannerbauer, H., Lehnert, M.D., Lutz, D., Tacconi, L.,
Bertoldi, F., Carilli, C., Ganzel, R., Menten, K., 2002, ApJ, 573, 473.

\bibitem{} Dunlop, J.S, et al., 2003, MNRAS, submitted.

\bibitem{} Fox, M.J., Efstathiou, A., Rowan-Robinson, M.,
Dunlop, J.S., Scott, S., Serjeant, S., et al., 2002, MNRAS, 331, 839.

\bibitem{} Frayer, D.T., Ivison, R.J., Scoville, N.Z., Yun, M., Evans,
A.S., Smail, I., Blain, A.W., Kneib, J.-P., 1998, ApJL, 506, L7.

\bibitem{} Frayer, D.T., Ivison, R.J., Scoville, N.Z., Evans, A.S.,
Yun, M., Smail, I., Barger, A.J., Blain, A.W., Kneib, J.-P., 1999,
ApJL, 514, L13.

\bibitem{} Gear, W.K., Lilly, S.J., Stevens, J.A., Clements, D.L.,
Webb, T.M., Eales, S.A., Dunne, L., 2000, MNRAS, 316, L51.

\bibitem{} Genzel, R., Baker, A.J., Tacconi, L.J., Lutz, D., 
Cox, P., Guilloteau, S., Omont, A., ApJ, 2002, in press.

\bibitem{} Genzel, R., et al., 2003, in prep.

\bibitem{} Goldader, J.D., Meurer, H., Heckman, T.M., Seibert, M.,
Sanders, D.B., Calzetti, D., Steidel, C.C., 2002, ApJ, 568, 651.

\bibitem{} Gonz\'alez Delgado, R.M., Heckman, T.M., Leitherer, C.,
Meurer, G., Krolik, J., Wilson, A.S., Kinney, A., Koratkar, A., 1998,
ApJ, 505, 174.

\bibitem{} Graham, J.R., Liu, M.C., 1995, ApJ, 449, L29.

\bibitem{} Heckman, T.M., Armus, L., Miley, G.K., 1990, ApJS, 74, 833.

\bibitem{} Heckman, T.M., Rovert, C., Leitherer, C., Garnett, D.R.,
van der Rydt, F., 1998, ApJ, 503, 646.

\bibitem{} Ibata, R.A., Lewis, G.F., Irwin, M.J., Leh\'ar, J., Totten,
E.J., 1999, AJ, 118, 1922.

\bibitem{} Ivison, R.J., Smail, I., Le Borgne, J.-F., Blain, A.W., Kneib,
J.-P., B\'ezecourt, J., Kerr, T.H., Davies, J.K., 1998, MNRAS, 298, 583.

\bibitem{} Ivison, R.J., Smail, I., Barger, A., Kneib, J.-P., Blain, A.\ W.,
Owen, F.N., Kerr, T.H., Cowie, L.L., 2000, MNRAS, 315, 209.

\bibitem{} Ivison, R.J., Smail, I., Frayer, D.T., Kneib, J.-P., Blain, A.W.,
2001, ApJL, 561, L45.

\bibitem{} Ivison, R.J., Greve, T.R., Smail, I., Dunlop, J.S., Roche,
N.D., Scott, S.E., et al., 2002, MNRAS, 337, 1.

\bibitem{} Iwamuro, F., Motohara, K., Maihara, T., Hata, R., Harashima, T.,
2001, PASJ, 53, 355.

\bibitem{} Ledlow, M., Smail, I., Owen, F.N., Keel, W.C., Ivison, R.J.,
Morrison, G.E.\ 2002, ApJ, 577, L79.

\bibitem{} Leitherer, C., Schaerer, D., Goldader, J.D., Gonz\'alez
Delgado, R.M., Carmelle, R., Kune, D.F., de Mello, D.F., Devost, D.,
Heckman, T., 1999, ApJS, 123, 3.

\bibitem{} Meurer, G.R., Heckman, T.M., Lehnert, M.D., Leitherer, C.,
Lowenthal, J., 1997, AJ, 114, 54.

\bibitem{} Meurer, G.R., Heckman, T.M., Calzetti, D., 1999, ApJ, 554, 1021.

\bibitem{} Osterbrock, D.E., 1981, ApJ, 249, 462.

\bibitem{} Page, M.J., Stevens, J.A., Mittaz, J.P.D., Carrera, F.J.,
2001, Science, 294, 2516.
 
\bibitem{} Pettini, M., Rix, S.A., Steidel, C.S., Adelberger, K.L.,
Hunt, M.P., Shapley, A.E., 2002, ApJ, 569, 742.

\bibitem{} Rengarajan, T.N., Takeuchi, T.T., 2001, PASJ, 53, 433.

\bibitem{} Schlegel, D., Finkbeiner, D.P., Davis, M., 1998, ApJ, 500, 525.

\bibitem{} Scott S.E., Fox M.J., Dunlop J.S., Serjeant S., Peacock
J.A., et al.\ 2002, MNRAS, 331, 817.

\bibitem{} Shapley, A.E., Steidel, C.C., Pettini, M., Adelberger,
K.L., 2003, ApJ, in press.

\bibitem{} Shu, C., Mo, H.J., Mao, S., 2003, MNRAS, submitted.

\bibitem{} Smail, I., Ivison, R.J., Blain, A.W., Kneib, J.-P., 2002,
MNRAS, 331, 495.

\bibitem{} Smail, I., Ivison, R.J., Gilbank, D.G., Dunlop, J.S., Keel,
W.C., Motohara, K., Stevens, J.A., 2003, ApJ, in press.

\bibitem{} Solomon, P., Downes, D., 2002, ApJ, in press.

\bibitem{} Soucail, G., Kneib, J.-P., B\'ezecourt, J., Metcalfe, L.,
Altieri, B., Le~Borgne, J.-F., 1999, A\&A, 343, L70.

\bibitem{} Strickland, D.K., Heckman, T.M., Weaver, K.A., Hoopes, C.G,
Dahlem, M., 2002, ApJ, 568, 689.

\bibitem{} Vernet, J., Cimatti, A., 2001, A\&A, 380, 409.

\bibitem{} Webb, T.M.A., Eales, S.A., Lilly, S.J., Clements, D.L.,
Dunne, L., Gear, W.K., Flores, H., Yun, M., 2002, ApJ, in press.

\bibitem{} Williams, L.L.R., Lewis, G.F., 1996, MNRAS, 281, L35.

\end{thebibliography}
\end{document}